\documentclass[twocolumn,amsmath,amssymb,prl,10pt,nofootinbib,superscriptaddress]{revtex4}
\usepackage{ graphicx, float,amsmath, amsmath, amssymb}
\usepackage[export]{adjustbox}

\def\be{\begin{equation}}
\def\ee{\end{equation}}
\def\bea{\begin{eqnarray}}
\def\eea{\end{eqnarray}}
\def\bse{\begin{subequations}}
\def\ese{\end{subequations}}

\usepackage[breaklinks, colorlinks, citecolor=blue]{hyperref}
\usepackage[labelsep=period]{caption}
\begin{document}
\title{Bouncing black holes in quantum gravity and the Fermi gamma-ray excess}

\author{Aur\'elien Barrau}
\email{Aurelien.Barrau@cern.ch}
\affiliation{
Laboratoire de Physique Subatomique et de Cosmologie, Universit\'e Grenoble-Alpes, CNRS-IN2P3\\
53,avenue des Martyrs, 38026 Grenoble cedex, France\\
}%

\author{Boris Bolliet}
\email{Bolliet@lpsc.in2p3.fr}
\affiliation{
Laboratoire de Physique Subatomique et de Cosmologie, Universit\'e Grenoble-Alpes, CNRS-IN2P3\\
53,avenue des Martyrs, 38026 Grenoble cedex, France\\
}%

\author{Marrit Schutten}
\email{M.Shutten@students.ru.nl}
\affiliation{
Laboratoire de Physique Subatomique et de Cosmologie, Universit\'e Grenoble-Alpes, CNRS-IN2P3\\
53,avenue des Martyrs, 38026 Grenoble cedex, France\\
}%
\affiliation{
Radboud University, Institute for Mathematics, Astrophysics and Particle Physics,\\
Mailbox 79, P.O. Box 9010, 6500 GL Nijmegen, The Netherlands.\\
}%

\author{Francesca Vidotto}
\email{F.Vidotto@science.ru.nl}
\affiliation{
Radboud University, Institute for Mathematics, Astrophysics and Particle Physics,\\
Mailbox 79, P.O. Box 9010, 6500 GL Nijmegen, The Netherlands.\\
}%

\date{\today}
\begin{abstract} 
Non-perturbative quantum-gravity effects can change the fate of black holes and make them bounce 
in a time scale shorter than the Hawking evaporation time.
In this article, we show that this hypothesis can account for the GeV excess observed from the galactic center by the {\it Fermi} satellite. By carefully taking into account the secondary component due to the decay of unstable hadrons, we show that the model is fully self-consistent. This phenomenon presents a specific redshift-dependance that could allow to distinguish it from other astrophysical phenomena possibly contributing to the GeV  excess. 
 \end{abstract}
\maketitle
\section{Introduction}

The Planck scale is currently out of reach from any direct local experiment by a factor of approximately $10^{15}$. It is therefore hard to test quantum gravity. Many efforts have however been devoted to quantum gravity phenomenology in the last decade (see, {\it e.g.}, \cite{Liberati:2011bp,Hossenfelder:2009nu,AmelinoCamelia:2008qg} and references therein for some general arguments) and it is not unreasonable to expect measurable consequences. Most efforts in the recent years have focused on the early Universe or on modified dispersion relations impacting the propagation of gamma-rays on huge distances. 
In this article, we focus on a recent result associated with black holes physics, first exposed in \cite{Rovelli:2014cta}. The main idea is grounded in a robust result of loop quantum cosmology: quantum gravity might manifest itself in the form of an effective pressure that counterbalances the classically attractive gravity when matter reaches the Planck density \cite{Ashtekar:2006rx}. For a black hole, this means that matter's collapse could stop before the central singularity forms. The classical singularity is replaced in the quantum theory by a phase of maximum density -- a ``Planck star" \cite{Rovelli:2014cta}. The absence of the central singularity allows for the dynamical trapping horizon (shrinking of light surfaces) 
to be converted in an anti-trapping horizon  (expanding of light surfaces), that releases matter and eventually disappears. 
This is a non-pertubartive quantum-gravity process that tunnels a classical black hole into a classical white hole.
Because of the gravitational redshift, the process is almost instantaneous in proper time but appears as very long if measured by an external distant observer.  

The viability of the model is supported by the existence of a classical metric satisfying the Einstein equations outside the spacetime region where matter collapses into a black hole and then emerges from a white hole\footnote{A modifications was suggested in \cite{DeLorenzo:2015gtx} where the scenario was made asymmetric, with a black hole phase longer than the white hole one. Such a modification overcomes complications coming from a possible instability in the white-hole phase.} \cite{Haggard:2014rza}. This can be achieved without violating causality nor the semiclassical approximation, as quantum effects piles up outside the horizon over a very long time. %

The time quantum effects take to pile up outside the horizon determines the lifetime of the black hole, and its phenomenology. This was first investigated in \cite{Barrau:2014hda} for a long lifetime (comparable but shorter than the Hawking evaporation time). Further studies in \cite{Barrau:2014yka} and \cite{Barrau:2015uca} were developed considering a wider range of possible lifetimes and the integrated signal coming from a diffuse emission. 

The tunneling process connects two classically disconnected solutions. Einstein equations should therefore be violated during the evolution, but the model allows for a violation that takes place only over a finite region. This is where full quantum gravity dominates\footnote{A possibility could be to study an effective metric associated with this finite region, as originally done by Hayward \cite{Hayward:2005gi}. See \cite{Lorenzo:2014hl} for recent results in this direction, recently extended to rotating metrics \cite{Lorenzo:2015zr}.}.
This process seems to be quite generically allowed for a wide range of viable quantum theories of gravity. Interestingly, in  covariant loop quantum gravity (LQG) it is possible to perform the calculation of the tunneling amplitudes  \cite{Christodoulou:2016vny} that provides an estimation of the black-hole lifetime.\\

In this work, we address the puzzle posed by the observation by the {\it Fermi} telescope of a GeV  photon excess, coming from the galactic center. Different explanations -- including standard astrophysical sources -- have been considered to explain it. 
Here we investigate whether bouncing (primordial) black holes could explain this specific excess and if this hypothesis has specific features that could allow to distinguish it from more conventional explanations.\\


In the first part, we briefly explain what are the parameters of the model and their possible values. In the second part, we present the way we have calculated and modeled the gamma-ray emission from bouncing black holes. In the third part we show the fit to the GeV {\it Fermi} excess we are interested in. In the fourth part, we suggest ways to discriminate our model from other possible explanations and normalize the masse spectrum. Some prospects are then discussed in the conclusion.

\section{Parameters of the model}
A precise astrophysical model for the emission from a bouncing black hole is not available, but heuristic arguments lead to consider  two different emission mechanisms \cite{Barrau:2014yka}. 
One, designated as the \emph{low-energy} component, is grounded in a simple and conservative dimensional analysis. The mean energy of the emitted signal is assumed to be such that the corresponding wavelength matches the size of the horizon. This is a reasonable expectation, agreeing with the Hawking spectrum. The other one, designated as the \emph{high-energy} component, has a smaller wavelength and depends on the conditions at which the black hole formed. In the model, the matter forming the black hole reemerges rapidly in the white-hole phase. The  gravitational blueshift felt by radiation in the contracting phase is precisely compensated by the very same amount of redshift in the expanding phase. \\

If the considered model is correct, the bounce should take place for all kinds of black holes, but observable effects become experimentally accessible only for primordial black holes (PBHs), {\it i.e.} black holes that formed in the early universe with a potentially wide mass spectrum. In particular, they can form with with masses smaller than the Solar mass so that their bouncing time can be of the order of the age of the Universe (more massive black holes would require much more than the Hubble time to bounce and nothing would be visible). Studying the phenomenology of bouncing black holes, we are  interested only by primordial black holes. Many different processes that can lead to the formation of black holes in the early Universe were suggested, see, {\it e.g.}, \cite{Green:2014qf} for a recent review. In the simplest models, PBHs form by collapse of over-dense regions. Given the mass of a black hole, its formation time is then (approximately) known and so is the spectrum of the radiation that collapsed to form it -- and that will emerge from the bounce in the \emph{high-energy} component of the signal considered here.\\

The most important parameter of the model is the bouncing time of black holes. It can be written as \cite{Haggard:2014rza} 
\begin{equation}\label{bt}
\tau = 4kM^2,
\end{equation}
in Planck units, where $M$ is the mass of the black hole and $k$ is a free parameter. This is a key-point: the bounce time scales as $M^2$ whereas the Hawking evaporation requires a time of order $M^3$. The parameter $k$ is bounded from below at the value $k_{min} = 0.05$ which ensures that the quantum effects do pile up enough to appear outside of the black hole horizon so that the bounce can take place. It is also bounded from above at a value $k_{max}(M)$ which translates the fact that the bouncing time needs to be smaller than the Hawking time\footnote{More precisely, the bounce time is constrained to be smaller than ``Page time'' at which the black holes would have lost half of its mass by Hawking evaporation because this time signs the entrance in the full quantum gravity regime \cite{ Almheiri:2012rt}.}, otherwise the black hole would disappear before bouncing and the evaporation could not be considered anymore as a small correction associated with a dissipative process, as assumed in the model.\\

A signal detected today comes from black holes that have lived for a time equal to the Hubble time $t_H$. Fixing the lifetime to $t_H$, Eq. \eqref{bt} gives the corresponding mass of the bouncing black hole, that determines the energy of the emitted radiation. We ask the following question: is there an allowed value of $k$  such that this emission can explain the GeV excess observed by the {\it Fermi} telescope? 
We note immediately that the GeV energy scale is far below any possible contribution coming from the high-energy component of our model: even for the smaller possible value of $k$ the emitted energy is of order a TeV. On the other hand, the low-energy component can indeed match the observed signal. Our analysis therefore focuses on this component.
To have an emitted energy of the order of 1 GeV, that is of order $10^{-19} E_{Pl}$, the size of the black hole should be of the order of $10^{19} l_{Pl}$ and its mass of the oder $M\sim 10^{19} M_{Pl}$. The Hubble time is $t_H\sim 10^{60} t_{Pl}$. Requiring the Hubble time to be equal to the bouncing time leads to $k\sim 10^{22}$. How does this compare with the Hawking time? The Hawking time is roughly $t_{Haw}\sim10^3M^3$, that is of the order of $10^{60} t_{Pl}$ for the mass we are interested in. This is of the same order of magnitude than the bouncing time\footnote{In our study we disregard the mass loss due to Hawking evaporation. In fact, even if the bouncing time considered here is comparable with the Hawking one, Hawking radiation decreases the mass of the black hole only by a small amount without changing its order of magnitude.}. This is therefore a quite interesting situation from the theoretical point of view in the sense that the required value of the parameter is not random or arbitrary in the (very large) allowed interval but a near-extremal one.\\

To summarize, the {\it high-energy} component of the signal emitted by bouncing black holes cannot explain the {\it Fermi} excess but the {\it low-energy} component might do so if the free parameter $k$ is chosen near its highest possible value.

\section{Modeling of the gamma-ray emission}

Whatever the details of the emission mechanism, as soon as fundamental particles are emitted at energies higher than the QCD confinement scale, quarks and gluons are emitted and do fragmentate into subsequent hadrons. For a bouncing black hole emitting quanta with energies greater than, say, 100 MeV, it is required to consider not only the primary ({\it i.e.} direct) emission of gamma-rays but also the secondary component, due to the decay of unstable hadrons produced by fragmentation. This has been studied with analytical approximations for evaporating black holes in \cite{MacGibbon:1990zk,MacGibbon:1991tj}. In this work we use a full Monte Carlo analysis based the ``Lund" PYTHIA code (with some scaling approximations in the low energy range) \cite{Sjostrand:2014zea} to determine the normalized differential fragmentation functions $dg(\epsilon,E)/d\epsilon$, where $E$ is the quark energy and $\epsilon$ is the photon energy. This takes into account a large number of physical aspects, including hard and soft interactions, parton distributions, initial- and final-state parton showers, multiple interactions, fragmentation and decay. \\

For all energies, we have found that the obtained spectra can be well fitted by a function 
\begin{equation}
f(E,\epsilon)=\frac{a\epsilon^b}{\pi\gamma}\left[\frac{\gamma^2}{(\epsilon-\epsilon_0)^2+\gamma^2}\right] e^{-\left(\frac{4\epsilon}{E}\right)^3},
\end{equation}
with $a=50.7$, $b=0.847$, $\gamma=0.0876$ and $\epsilon_0=0.0418$ if the energies are given in GeV. The low-energy peak of the spectrum is well approximated by a Cauchy function. It is then roughly a power law, followed by an exponential cutoff around the initial jet energy.

\begin{figure}[H]
\includegraphics[width=90mm,center]{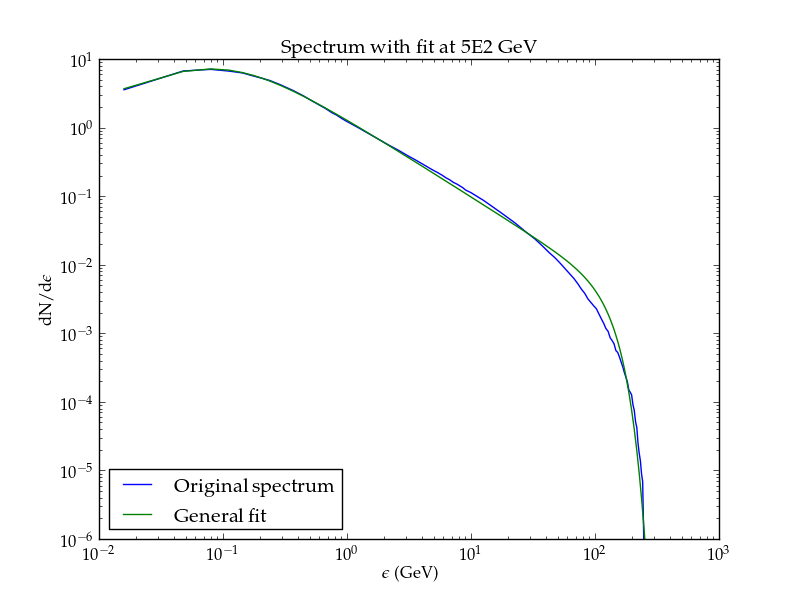}
\caption{Spectrum of gamma-rays generated by $5\times 10^2$ GeV jets. The green histogram corresponds to the output of the simulation and the blue curve to the analytical fit.}
\label{error}
\end{figure}

\begin{figure}[H]
\includegraphics[width=90mm,center]{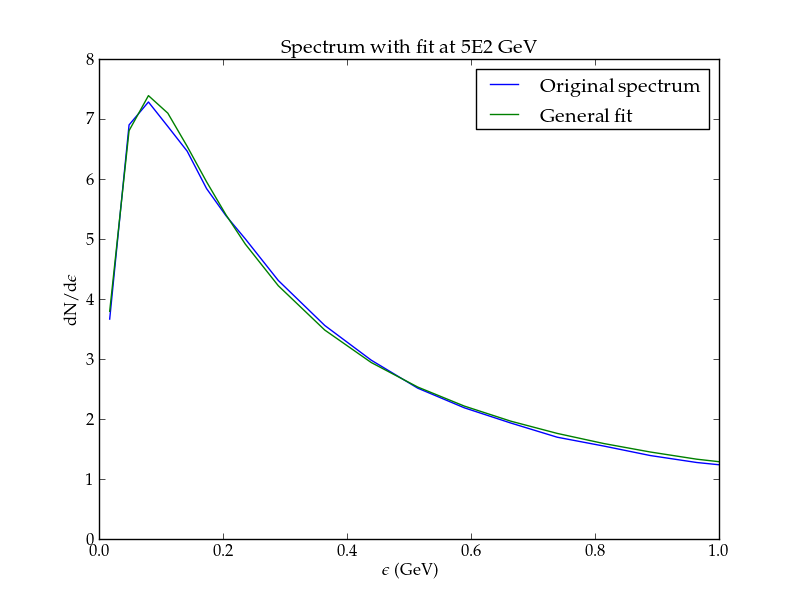}
\caption{Zoom on the low-energy part of the spectrum of gamma-rays generated by $5\times 10^2$ GeV jets. The green histogram corresponds to the output of the simulation and the blue curve to the analytical fit.}
\label{error}
\end{figure}

As soon as the jet reaches an energy much higher that the associated quark mass, the result does not depend substantially on the quark type. Depending on the mean energy $E$ of the primary component, the number of types of emitted quarks -- that is with $m<E$ -- is accounted for. The normalisation is chosen to be consistent with the primary emission. 

For the {\it low-energy} component, the shape of the primary signal is not completely determined by the model. 
We have used a Gaussian function, centered on the energy estimated in the previous Section,
\color{black}
 with a relative width taken as the second free parameter of the model. Its exact value depends on the details of the astrophysical phenomena occurring during the bounce and this is far beyond the scope of this study. The full signal can be written as 
\begin{equation}
Ae^{-\frac{(\epsilon-E)^2}{2\sigma^2}}+3N\sqrt{2\pi}A\sigma f(E,\epsilon),
\end{equation}
where $N$ is the number of species of quarks with $m<E$.

For the {\it high-energy} component, which is irrelevant for this study but potentially interesting for other works, the same strategy can be followed. The primary component is then a Planck law and the full signal can be written as
\begin{equation}
A\frac{\epsilon^2}{e^{E/T}-1}+36AT^3\zeta(3) f(E,\epsilon).
\end{equation}
Interestingly, this  formula can also be used to model the full spectrum of an evaporating black hole since the Hawking spectrum is also very close to a Planck law.

\section{Fitting Fermi data}

The {\it Fermi} Gamma-ray Space Telescope is a space observatory being used for gamma-ray astronomy observations from low Earth orbit. Its main instruments are the Large Area Telescope (LAT), intended to perform an all-sky survey studying astrophysical and cosmological phenomena, and the Gamma-ray Burst Monitor (GBM), used to study transients.

An excess in the {\it Fermi}-LAT data has been reported within the inner 10 arcmin of the Galactic center (see, {\it e.g.}, \cite{Hooper:2010mq,Abazajian:2012pn,Gordon:2013vta}) and up to larger galactic latitudes (see, {\it e.g.}, \cite{Daylan:2014rsa,Calore:2014xka,Hooper:2013rwa,Huang:2013pda}). A huge number of works have been published on possible explanations.  Our opinion is that an astrophysical origin, notably associated with millisecond pulsars, is the most convincing one (see, {\it e.g.}, \cite{Bartels:2015aea}). It is however not fully satisfactory and dark-matter like hypotheses are worth being considered (see, {\it e.g.}, \cite{Daylan:2014rsa}). Here we investigate whether this signal can be due to bouncing black holes.\\

\begin{figure}[H]
\includegraphics[width=90mm,center]{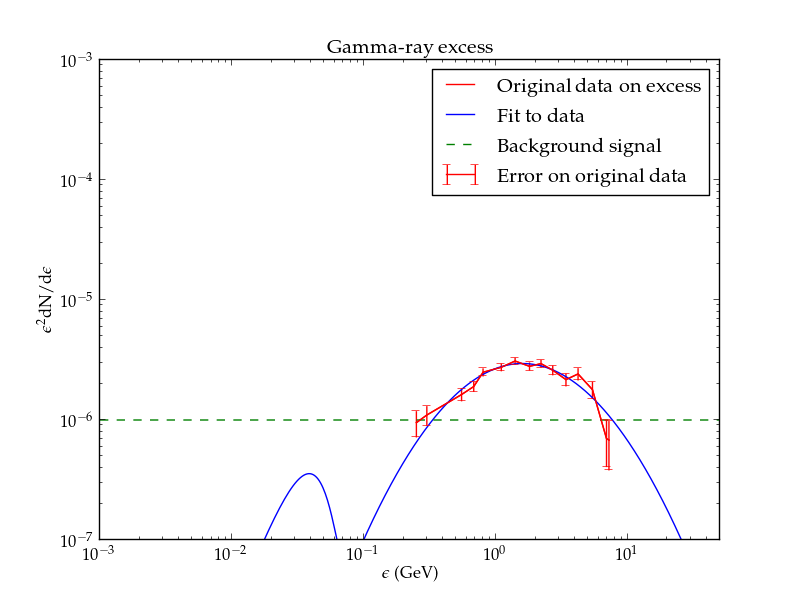}
\caption{Best fit to the {\it Fermi} excess with bouncing black holes.}
\label{fit}
\end{figure}

We stress that the explanation we suggest is specifically associated with the quantum gravity scenario considered in this work. The time integrated spectrum of black holes evaporating by the usual Hawking process is scaling as $E^{-3}$ and there is no way it can account for the {\it Fermi} excess. As explained before, two  parameters are required to fully determine the low-energy component of bouncing black holes: their bouncing time and the width of the primary Gaussian. The best fit (with a near-extremal bouncing time) is shown in Fig. \ref{fit}. The agreement with data is good, with a $\chi^2$ per degree of freedom of $1.05$. Notice that what is plotted here is not the differential spectrum but the spectral energy density $(\epsilon^2dN/d\epsilon)$, as used for most experimental publications. The key point we want to stress is that although the number of secondary gamma-rays is higher than the number of primary gamma-rays, their spectral energy density is much lower. This is of utmost importance for this study: as the background has a basically constant spectral energy density, this means that the anomaly can be accounted for without any spurious excess in the 10-100 MeV range, where is situated the peak of the secondary component. This peak remains much below the background and the signal can be explained with no contradiction with the data.

This also shows why the {\it high-energy} component cannot be used to explain the excess. The energy of its primary component is in all cases too high and its secondary component would not have a high enough spectral energy density.

\section{Discrimination with dark matter and mass spectrum}

\begin{figure}[H]
\includegraphics[width=85mm,center]{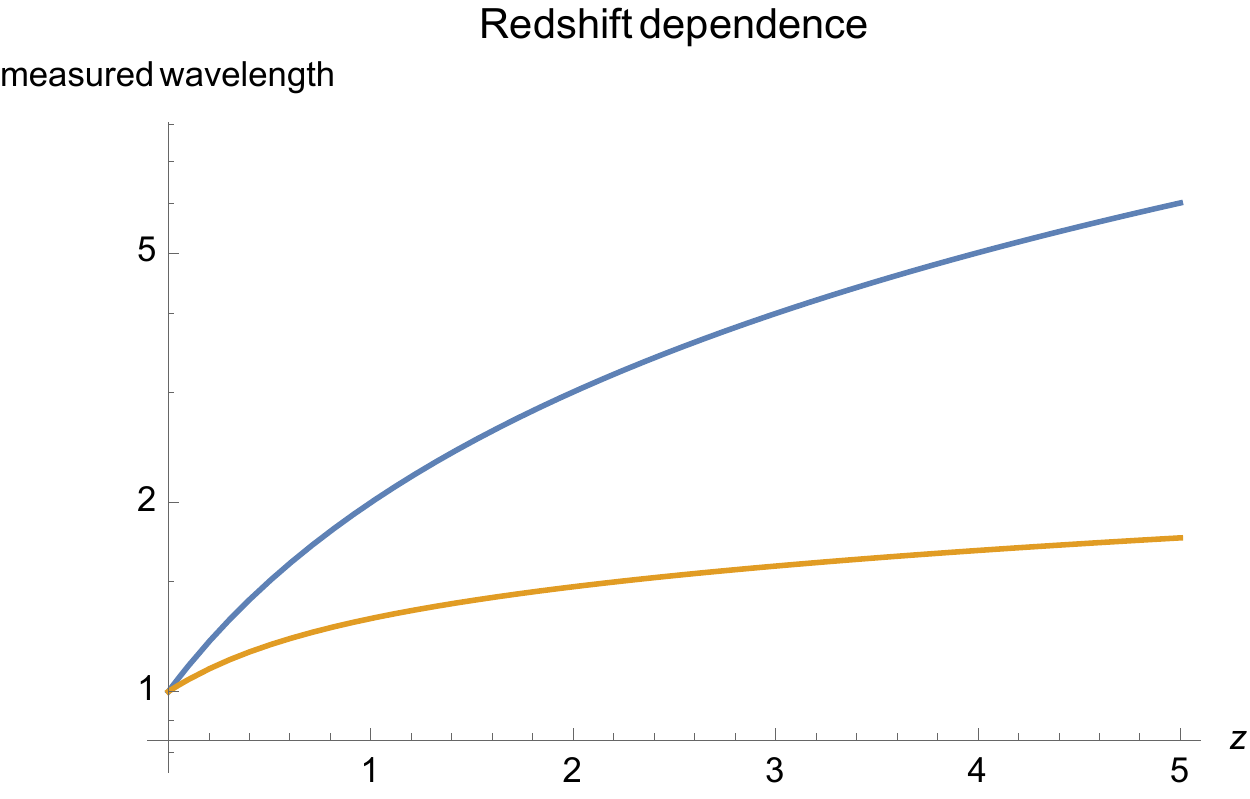}
\caption{Measured wavelength, normalized to the rest-frame wavelength, as a function of the redshift. The upper curve is for a conventional signal and the lower curve is for the model considered in this article.}
\label{red}
\end{figure}

The model presented in this work is unquestionably quite exotic when compared with astrophysical hypotheses. But the important point is than it can, in principle, be distinguished both from astrophysical explanations and from other ``beyond the Standard Model" scenarios. The reason for that is a peculiar redshift dependance. When looking at a galaxy at redshift $z$, the measured energy of the signal emitted either by decaying WIMPS or by astrophysical objects will be $E/(1+z)$ if the rest-frame energy is $E$. But this is not true for the bouncing black holes signal. The reason for this is that black holes that have bounced far away and are observed now must have a shorter bouncing time and therefore a smaller mass. Their emission energy -- in the low energy channel we are considering in this article -- is therefore higher and this partly compensates for the redshift effect. Following \cite{Barrau:2014yka}, we can write down the observed wavelength of the signal from a host galaxy at redshift $z$, taking into account both the expansion of the universe and the change of bouncing time, as:
\begin{eqnarray}
\lambda_{obs}^{BH}\!&\sim&\! \frac{2Gm}{c^2} (1+z)\ \ \times \\ \nonumber
&& \  \sqrt{\frac{H_0^{-1}}{6\,k\Omega_\Lambda^{\,1/2}
}\ \sinh^{-1}\!\!\left[ \left(\frac{\Omega_\Lambda}{\Omega_M}\right)^{\!1/2} (z + 1)^{-3/2}\right]},
\label{redhisft}
\end{eqnarray}
where we have reinserted the Newton constant $G$ and the speed of light $c$; $H_0, \Omega_\Lambda$ and $\Omega_M$ being the Hubble constant,  the cosmological constant, and the matter density. On the other hand, 
for standard sources, the measured wavelength is just related to the observed wavelength by
\begin{equation}
\lambda_{obs}^{other}=(1+z)\lambda_{emitted}^{other}~.
\end{equation}
The redshift dependence specific of our model makes it possibly testable against other proposals. Obviously, detecting such a signal from far away galaxies is challenging but we hope this work might 
 motivate some experimental prospects for the next generation of gamma-ray satellites. On Fig. \ref{red}, we have displayed the evolution of the wavelength, normalized to the rest-frame wavelength, as a function of the redshift for both a conventional source (upper curve) and the model considered in this work (lower curve). By ``conventional" we mean here basically all other models we are aware of, including astrophysical sources and the decay of supersymmetric particles.  Obviously it is easy to distinguish between both cases: in the hypothesis of bouncing black holes, the wavelength does not vary much because black holes bouncing far away are smaller and therefore emit higher-energy photons.

Interestingly, there might be another specific observational signature for this model. In addition to specific signals coming from identified galaxies, one should also expect a diffuse background. As we have demonstrated in \cite{Barrau:2015uca}, for the {\it low energy} component of the bouncing signal, considered here, the integrated emission exhibits an interesting feature. The integrated spectrum, defined as 
\begin{equation}
\frac{dN_{mes}}{dEdtdS}=\int\Phi_{ind}((1+z)E,R) \cdot n(R) \cdot A(E) \cdot f(E,R)  dR,
\label{flux_int}
\end{equation}
(where $\Phi_{ind}(E,R)$ denotes the individual flux emitted by a single bouncing black hole at distance $R$ and at energy $E$, $A(E)$ is the angular acceptance of the detector multiplied by its efficiency, $f(E,R)$ is the absorption function, and $n(R)$ is the number of black holes bouncing at distance $R$ per unit time and volume) was indeed shown to be nearly the {\it same} than the individual spectrum but with a slight distorsion on the left tail \cite{Barrau:2015uca}. This is another signature for this specific model.\\

The order of magnitude of the number of bouncing black holes in the galactic-center region required to account for the observed flux is around $100$ per second. The associated 
mass is  negligible when compared to the expected dark matter density, even when integrated over a long time interval. If the mass spectrum of primordial black holes was known, which is not the case, in principle it would be possible to fix the total mass associated with bouncing black holes. As a reasonable toy model, let us assume that the mass spectrum is given by 
\begin{equation}
\frac{d^2N}{dMdV}=pM^{-\alpha}.
\end{equation}
If the number of exploding black holes required to explain the data on a time interval $d\tau$ is $N_{exp}$, one can estimate the associated mass variation
\begin{equation}
dM=\frac{d\tau}{8kM}.
\end{equation}
Calling $M_0$ the mass corresponding to a black hole exploding today, one then gets
\begin{equation}
N_{exp}=\int_{M_0}^{M_0+dM}pM^{-\alpha}dM\,.
\end{equation}
This allows, in principe, to determine $p$ and therefore to normalize the spectrum.

\section{Conclusion}

Black holes could bounce once they have reached the ``Planck star" stage. This can be seen as a tunneling into an expanding explosive phase. The process appears generic in  quantum gravity.  In this article, we have shown that this phenomenon could explain the GeV excess measured by the {\it Fermi} satellite. This would open the fascinating possibility to observe (non perturbative) quantum gravity processes at energies 19 orders of magnitude below the Planck scale. Interestingly, the explanation we suggest is fully self-consistent in the sense that the hadronic ``noise" due to decaying pions remains much below the observed background. Unquestionably, there are other -- less exotic -- ways to explain the {\it Fermi} excess. But the important point we have made is that this model has a specific redshift dependance which, in principle, can lead to a clear signature for future experiments. On the theoretical side, the important next step would be to fix the free parameter of the model from the full theory so that the energy of the signal is fixed from first principle and not anymore tuned to fit the data (see \cite{Christodoulou:2016ve} for a recent step in this direction). Another interesting possible improvement would be to take into account the distribution of actual bouncing times for individual black holes around the mean time $\tau$ fixed by the theory.

 \section{Acknowledgments}

B.B. is supported by a grant from ENS-Lyon.
F.V. is supported by a Veni grant from the Netherlands Organisation for Scientific Research (NWO).

\bibliography{refs,bib}
 \end{document}